\documentclass[letterpaper, 10pt, conference]{ieeeconf}     
\makeatletter
\let\NAT@parse\undefined
\makeatother
\pdfoutput=1

\usepackage{amsfonts}
\usepackage{amsmath}
\usepackage{amssymb}
\usepackage{color}
\usepackage[numbers]{natbib}
\usepackage{graphics}
\usepackage{caption}
\usepackage{epsfig}
\usepackage{chngcntr}
\usepackage{dashrule}
\usepackage[position=b]{subcaption}
\usepackage[hyperfootnotes=false]{hyperref}
\usepackage{listings}
\usepackage{color}
\usepackage{authblk}


\usepackage{amsfonts}
\usepackage{graphicx}

\usepackage{ifthen}
\usepackage{color}

\def\send#1#2{\stackrel{#1}{\hbox to #2{\rightarrowfill}}}
\def\-{\!\!\!\!\!-}

\newtheorem{theorem}{Theorem}
\newtheorem{lemma}{Lemma}

\newcounter{seqn}[equation]
\def\theseqn{\arabic{equation}\alph{seqn}}

\def\endseqn{\eqno \@seqnnum
$$\ignorespaces}
\def\@seqnnum{(\theseqn)}

\newskip\mcentering \mcentering=0pt plus 1000pt minus 1000pt

\def\meqalignno#1{
\halign to\displaywidth{
    \hbox to 0pt{\kern\displaywidth\llap{$##$}\hss}\tabskip=\mcentering
    &\hfil$\displaystyle{##}$\tabskip=\mcentering
   &&$\displaystyle{{}##}$\hfil\tabskip=\mcentering
    \crcr
    #1\crcr}}

\def\dspace{\multiply\normalbaselineskip 150
		  \divide\normalbaselineskip 100 \normalbaselines
		  \csname @@normalbaselineskip\endcsname\normalbaselineskip}
\def\sspace{\multiply\normalbaselineskip 200
		 \divide\normalbaselineskip 300 \normalbaselines
		 \csname @@normalbaselineskip\endcsname\normalbaselineskip}
\def\sdspace{\multiply\normalbaselineskip 160
		 \divide\normalbaselineskip 150 \normalbaselines
		 \csname @@normalbaselineskip\endcsname\normalbaselineskip}

\def\@{\tilde}

\def\3dot#1{\buildrel\textstyle...\over#1}

\IEEEoverridecommandlockouts                          

\title{\LARGE \bf
Game of Power Allocation on Networks}

\author{Yuke Li,\thanks{This work was supported by National Science Foundation grant n.1607101.00 and US Air Force grant n. FA9550-16-1-0290.} and A. S Morse\thanks{Y. Li and A. S. Morse are respectively with the Department of Political Science, and the Department of Electrical Engineering, Yale University, New Haven, CT, USA, \{yuke.li, as.morse\}@yale.edu}}

\begin{document}

\maketitle
\thispagestyle{empty}
\pagestyle{empty}

\begin{abstract}

This paper develops a distributed resource allocation game to study countries' pursuit of targets such as self-survival in the networked international environment. The contributions are two. First, the game formalizes countries' power allocation behaviors which fall into the broad category of humans resource allocation behaviors. Second, the game presents a new technical problem, and establishes pure strategy Nash equilibrium existence.


\end{abstract}

\section{Introduction}

\subsection{Question}


What fundamentally drives politics in international relations? This paper takes to position that the answer is the way in which each country's ``power'' is allocated where by the \emph{power} of a country is meant a concept adopted from \citep{morgenthau1973politics}. This concept suggests that national power aggregates all the conventional and nuclear war making capabilities mobilizable for use in international security, being composed of geography, natural resources(e.g., food, raw materials), industrial capacity, military preparedness (e.g., technology, leadership, quantity and quality of armed forces), population (e.g., distribution and trends), national character, national morale and quality of diplomacy and government~\citep{morgenthau1973politics}. The real-world contexts underlying the environment (e.g., the nature of countries' engagement) are crucial to evaluating power. Several important aspects include, for instance, whether the engagement is real conflict or peacetime escalation. In peacetime, a country's military power may include its manpower, land systems, air power, naval power, logistical and financial support, and so forth. In wars, a country's power includes all its \emph{mobilized} resources. Also, even the national morale would become relevant for assessing countries' power in wars --- even if two countries have the same amount of power, the mutiny and morale problems could mark the difference between victory and defeat after controlling for the other factors. 

\begin{figure}[h]
	\centering
	\includegraphics[width=0.95\linewidth]{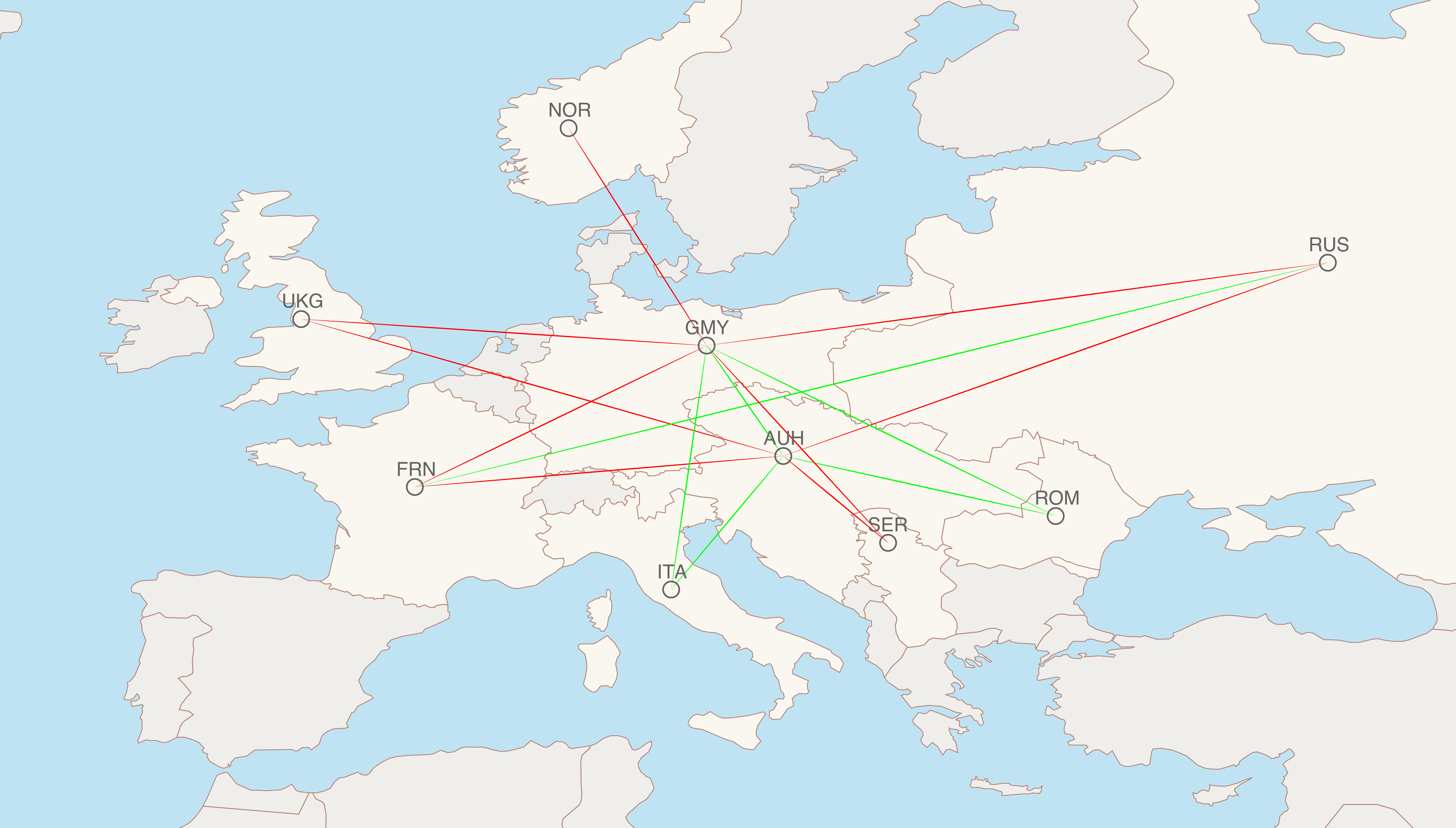}
	\caption{Red lines denote adversary relations and green lines denote friend relations.}
	\label{fig/1914}
\end{figure}

Power allocation is entirely different than other types of allocation (e.g., spectrum allocation, personnel allocation, etc) both in terms of its form and its objectives. We will be interested in a collection $\mathcal{C}$ of $n >0$ countries labelled from 1 to $n$. Distinct countries $i$ and $j$ are \emph{neighbors} if there is a ``bilateral relationship'' between the two which might be either a \emph{friend} relation or an \emph{adversarial} relation.  For instance, the collection of countries, $\{\texttt{GMY}, \texttt{UKG}, \texttt{RUS}, \texttt{FRN}, \texttt{AUH}, \texttt{ITA}, \texttt{ROM}, \texttt{SER}, \texttt{NOR}\}$, shown in the map of Europe in Figure~\ref{fig/1914}, were involved in WWI as of July 28, 1914, and labelled from 1 to 9. Specifically speaking, a country allocates its power to itself and every neighbor.  Though its power may increase over time, this kind of allocations happen all of the time. 

%
%

Power allocation theoretically generalizes many kinds of international affairs a country may possibly pursue, e.g., deploying arms and personnels to attack an adversary or to militarily assist a friend. By doing so, the country aims to support the survival of themselves and their friends and oppose that of their adversaries, and, amongst all, makes its self-survival the first priority, just as stated in the inaugural address of John F. Kennedy in 1961,  ``we shall pay any price, bear any burden, meet any hardship, support any friend, oppose any foe to assure the survival and the success of liberty''. This is also consistent with a basic assumption that has been held in the literature of international relations theory ever since the publication of \cite{waltz1979theory}, that is, countries primarily seek self-survival as the minimal target, and, when this target is satisfied, would pursue ``universal domination'' at a maximum target -- ``Universal domination'' can definitely be interpreted as a form of ``success'' mentioned above, which at least reflects in jeopardizing the adversaries' survival, and also, invariably, in guaranteeing the friends' survival.

A natural question arises: could a country always survive in an arbitrary international environment? To answer this question, we develop a general framework, which is a distributed resource allocation game on networks, for studying all possible directions of countries' power allocation in \emph{any} networked international environment, i.e., environments that have existed or still exist or that are hypothetical,  and the implications for the realization of countries' targets.

\subsection{Contributions}

Two features below about our \emph{power allocation game} are:

\begin{enumerate}
\item The game has an infinite strategy space -- as a country's power is a nonnegative real number, it always has infinite possibilities of allocating its power, which stands in contrast with a finite game where agents only need to choose from a finite set of options. 

\item  The game does not assume a particular group of countries, particular levels of total power and particular friend or adversary relations among them. Instead,  many possibilities for the set of countries, their total power and relation configurations are accounted for in the game. 


\end{enumerate}

The two features help to define a myriad of possibilities for the networked international environment and then enable a rich set of predictions.  To the best of our knowledge, a resource allocation game like this has never been studied before. Also, very few papers on similar problems exist, for which an example is \citep{jackson2015networks} that focuses on the problem of whether world trade helps to decrease the frequency of militarized conflicts. 

Broadly speaking, the distributed power allocation game can be used for studying two aspects below: 

\begin{enumerate} 

\item an understanding of countries' ``everyday politics''. Due to its many forms, a country's power allocation does not necessarily involve real conflicts but can be strategies to deter the adversaries and reassure the friends.

\item an assessment of the favorability of any given networked international environment for any country involved. For instance, a question that can be asked is, are there certain environments that may enhance a country's survivability? 

\end{enumerate}

\section{The Networked International Environment}

As will be consecutively defined, a ``networked international  environment'' consists of a set of countries, their total power, a set of ``admissible strategy matrices'', ``support and threat functions'' for country $i$, ``state space'', and a notion of ``preferences'' which we formalize with ``the axioms of choice''.

\subsection{Countries} As in the introduction, there is a collection $\mathcal{C}$ of $n$ countries, labeled $1,2, ..., n$; let the set of labels be $\mathbf{n} = \{1,2,3, ..., n\}$.



\subsection{Total Power} Each country $i$ in $\mathcal{C}$ has a given nonnegative quantity $p_{i} \in \mathbb{R}$ called its \emph{total power}. By a \emph{total power vector} for the countries is meant a real, nonnegative valued row vector $p = [p_{i}]_{1 \times n}$.


\subsection{Friend and Adversary Relations} Apart from itself, each country $i$ has  a subset of countries in $\mathcal{C}$ called \emph{friends} with labels in $\mathcal{F}_{i} \subset \mathbf{n}$ and a subset of countries in $\mathcal{C}$ called \emph{adversaries} with labels in $\mathcal{A}_{i} \subset \mathbf{n}$. We assume that for each $i \in \mathbf{n}$, $\{i\}, \mathcal{F}_{i}$, $\mathcal{A}_{i}$ are disjoint sets. We assume that these relations are symmetric, that is, if $j$ is a friend (or an adversary) of $i$, then $i$ is a friend (or an adversary) of $j$.  

The unordered pair $\{i,j\}$ stands for a pair of distinct labels in $\mathbf{n}$ such that $i$ and $j$ have a relation.  Denote the set of all friendly pairs as $\mathcal{R}_{f}$ and the set of all adversarial pairs as $\mathcal{R}_{a}$. Suppose the number of pairs in $\mathcal{R}_{f} \cup \mathcal{R}_{a}$ is $m$. A map $\eta: \mathcal{R}_{f} \cup \mathcal{R}_{a} \to \mathbf{m}$ where $\mathbf{m} = \{1,2, ..., m\}$ determines for each element in $\mathcal{R}_{f} \cup \mathcal{R}_{a}$ a distinct label in the set $\mathbf{m}$.


%

\subsection{Power Allocation Strategy} By country $i$'s \emph{power allocation strategy} is meant a real, nonnegative valued row vector $u_{i} \in \mathbb{R}^{1 \times n}$ whose $j$-th entry is $u_{ij}$. If $j \in \mathcal{F}_{i}$, then $u_{ij}$ represents the portion of country $i$'s total power which country $i$ is willing commit to the support or defense of friend $j$ against friend $j$'s adversaries. Similarly, if $j \in \mathcal{A}_{i}$, then $u_{ij}$ is the portion of country $i$'s total power that it is committing to its possible offense actions against country $j$. In addition, $u_{ii}$ is the portion of country $i$'s total power it holds in reserve. Finally, if $j \not\in \{i\} \cup \mathcal{F}_{i} \cup \mathcal{A}_{i}$, we stipulate that $u_{ij} = 0$.

Accordingly, for each $i \in \mathbf{n}$, $\sum_{j = 1}^{n}u_{ij} = p_{i}$ so the $i$-th row sum of the \emph{strategy matrix} $U = [u_{ij}]_{n \times n}$ is $p_{i}$. We write $\mathcal{U}$ for the set of all admissible strategy matrices. 


\subsection{Support and Threat} For each country $i \in \mathbf{n}$, we define two types of nonnegative-valued functions on $\mathcal{U}$.  The first, called a support function for country $i$ is the map $\sigma_{i}$: $\mathcal{U} \to [0, \infty)$, 
\begin{equation*}
U \longmapsto u_{ii} + \sum_{j\in \mathcal{F}_{i}}u_{ji} + \sum_{j \in \mathcal{A}_{i}}u_{ij}
\end{equation*} where for $j \in \mathcal{F}_{i}$, $u_{ji}$ is the total amount of power the friends of country $i$ commit to country $i$'s support or defense and for $j \in \mathcal{A}_{i}$, $u_{ij}$ is the total amount of power country $i$ commits to its possible offense against all of its adversaries. 

The second function, called a threat function for country $i$, is the map $\tau_{i}: \mathcal{U} \to [0, \infty)$, 

\begin{equation*}
U \longmapsto \sum_{j \in \mathcal{A}_{i}}u_{ji}
\end{equation*} Thus $\tau_{i}(U)$ is the total power all of country $i$'s adversaries commit to their respective offenses against country $i$. 

\subsection{State and State Space.} As a consequence of specific allocations, each country $i$ may find itself in one of three possible states, namely a \emph{safe} state, a \emph{precarious} state, or an \emph{unsafe} state. Let $x_{i} : \mathcal{U} \to \{\text{safe}, \text{precarious}, \text{unsafe}\}$ denote the \emph{state function}

\begin{equation*}
x_{i} \longmapsto \begin{cases}
\text{safe}   & \text{if}~ \sigma_{i}(U) > \tau_{i}(U)\\
\text{precarious}  & \text{if}~ \sigma_{i}(U) = \tau_{i}(U)\\
\text{unsafe} & \text{if}~ \sigma_{i}(U) < \tau_{i}(U)

\end{cases}
\end{equation*} We call $x_{i}(U)$ the state of country $i$ induced by power allocation strategy $U \in \mathcal{U}$. More generally, by the state of the overall collection of countries $\mathcal{C}$ induced by power allocation matrix $U$ is meant the row vector $x(U) = [x_{i}]_{1 \times n}$. The state space of $\mathcal{C}$ is thus the finite set $\mathcal{X} = \{x(U) : U \in \mathcal{U}\}$ whose cardinality is $3^{n}$.

\subsection{Axioms of Choice} We need to construct a rationale for each country to choose its own power allocation strategy. This will be done by comparing states for each country induced by different strategy matrices. We propose the following three axioms to capture the most intuitive aspects for any country's choice of its own power allocation strategy, which help to partially order the elements in $\mathcal{U}$. 


First, given country $j$,  call $V \in \mathcal{U}$ an ``admissible alternative'' to $U \in \mathcal{U}$ if $x_{k}(U) = x_{k}(V)$ for all $k \in \mathbf{n} \smallsetminus \{j\}$. We note that \emph{admissible alternative} is an equivalence relation on $\mathcal{U}$. From the perspective of country $i$, $U$ is \emph{weakly preferred} over $V$, i.e., $U$ is \emph{strictly preferred} over $V$ ($U \prec V$) or they are perceived indifferently ($U \sim V$), formally $U \preccurlyeq V$, if the respective condition in Axiom 1 holds for that particular country $j$, i.e, one of the following holds,

\subsubsection*{Axiom 1} 
\begin{itemize}
\item $j \in \{i\} \cup \mathcal{F}_{i}$,  $(x_{j}(V) \in \{\text{safe}, \text{precarious}\}) \lor (x_{j}(U) = \text{unsafe})$ 
\item $j \in \mathcal{A}_{i}$, $(x_{j}(V) \in \{\text{unsafe}, \text{precarious}\}) \lor (x_{j}(U) = \text{safe})$ 

\item $j \not\in \{i\} \cup \mathcal{F}_{i} \cup \mathcal{A}_{i}$, $x_{j}(U)$ and $x_{j}(V)$ may take any value \footnote{Here $U \preccurlyeq V$ trivially holds, and $U$ and $V$ are perceived indifferently, formally $U \sim V$.}

\end{itemize}

Second,  take an arbitrary element $V \in \mathcal{U}$ without requiring it to be an admissible alternative to $U$ as before. From the perspective of country $i$, $U$ is weakly preferred over $V$, formally $U \preccurlyeq V$, if the respective condition in Axiom 2 holds for any country $j \in \mathbf{n}$, i.e., all of the following holds,

\subsubsection*{Axiom 2} 

\begin{itemize}
\item $j \in \{i\} \cup \mathcal{F}_{i}$, $(x_{j}(V) \in \{\text{safe}, \text{precarious}\}) \lor (x_{j}(U) = \text{unsafe})$ 
\item $j \in \mathcal{A}_{i}$, $(x_{j}(V) \in \{\text{unsafe}, \text{precarious}\}) \lor (x_{j}(U) = \text{safe})$ 
\item $j \not\in \{i\} \cup \mathcal{F}_{i} \cup \mathcal{A}_{i}$, $x_{j}(U)$ and $x_{j}(V)$ may take any value

\end{itemize}

Third, we would like an axiom to capture countries' \emph{first priority of self-survival}, and take an arbitrary $V \in \mathcal{U}$ again. From the perspective of country $i$, $U$ is strictly preferred over $V$, formally $U \prec V$, if the below holds for itself, 

\subsubsection*{Axiom 3} 

\begin{itemize}
\item If $(x_{i}(V) \in \{\text{safe}, \text{precarious}\}) \land (x_{i}(U) = \text{unsafe})$ 

\end{itemize}In particular, Axiom 3 emphasizes that the best possible strategy matrix (or matrices) for country $i$ must be one(s) in which $i$ has achieved self-survival.

\subsection{Graphical Representation} 

A \emph{networked international environment} can now be formally defined as the collection of all the aforementioned elements or \emph{parameters}, $\{\mathcal{C},  p, \mathcal{U}, \sigma_{i}, \tau_{i}, \mathcal{X}, \preccurlyeq\}$. A directed and connected\footnote{The connected graph is assumed only for simplicity. For unconnected graphs, we only need to analyze the connected components separately.} graph on $n$ vertices and $2m$ edges, $\mathbb{G} = (\mathcal{V}, \mathcal{E})$, is a convenient representation of $n$ countries, $m$ pairs of which have a relation ($n, m \in \mathbb{Z}$), in an environment:

\begin{enumerate}
\item $\mathbb{G}$'s vertex set $\mathcal{V} = \{1, 2, \cdots, n\}$ is equivalent to the set of country labels, $\mathbf{n}$.

\item $\mathbb{G}$'s edge set $\mathcal{E} = \{(i,j): (i,j) = \eta(\{i,j\}), \{i,j\} \in \mathcal{R}_{a} \cup \mathcal{R}_{f}\}$, where $\eta$ maps every element $\{i,j\} \in \mathcal{R}_{a} \cup \mathcal{R}_{f}$ to two ordered pair $(i,j)$ and $(j,i)$ that represent the two edges with opposite directions between $i$ and $j$. Obviously, $|\mathcal{E}| = 2m$. 
\end{enumerate}

The weighted version of the above graph $\mathbb{G} = (\mathcal{V}, \mathcal{E})$ can represent a strategy matrix:

\begin{enumerate}
\item $u_{ij}$, which represents $i$'s allocations on the relation with $j \neq i$, is drawn as the edge weight of $(i,j)$. 
\item $u_{ii}$, which represents $i$'s power held in reserve, is drawn as the vertex weight of $i$.
\end{enumerate}

\section{Power Allocation Game}

\subsection{The Game}

The game of interest can be termed as a \emph{power allocation game} (PAG), which can be defined simply using the collection of elements denoting the networked international environment, $\Gamma = \{\mathcal{C},  p, \mathcal{U}, \sigma_{i}, \tau_{i}, \mathcal{X}, \preccurlyeq\}$.The power allocation game is \emph{static}, where each country chooses its own power allocation strategy at the same time.  

A \emph{complete information} framework is assumed, where countries have full knowledge of the environment, and this framework is also suitable for studying the scenarios where countries act upon individual and subjective perceptions of the environment.

\subsection{Equilibrium Concept}

 Let country $i$'s deviation from the strategy matrix $U$ be a nonnegative-valued $1\times n$ row vector $d_{i} \in \mathbb{R}^{1 \times n}$ such that $u_{i} + d_{i}$ is a valid strategy that satisfies the total power constraint. The deviation set $\mathcal{\delta}_{i}(U)$ is the set of all possible deviations of country $i$ from the strategy matrix $U$. A strategy matrix $U$ is a pure strategy Nash Equilibrium if no unilateral deviation in strategy by any single country $i$ is profitable for $i$ that is
\begin{equation*}U + e_{i} d_{i} \preccurlyeq U, \forall d_{i} \in \mathcal{\delta}_{i}(U)\end{equation*} where $e_{i}$ is an $n \times 1$ unit vector whose elements are $0$ but the $i$-th coordinate which is $1$.

\subsection{Equilibrium Class}

Denote the set of pure strategy Nash equilibria as $\mathcal{U}^{*}$\footnote{As will be proved in E, it will be an nonempty set.}. Call $V^{*}$ an admissible alternative to $U^{*} \in \mathcal{U}^{*}$ if $x(U^{*}) = x(V^{*})$.  Let $[\mathcal{U}^{*}]_{x(U^{*})}$ be the \emph{equilibrium equivalence class} of $U^{*} \in \mathcal{U}^{*}$. Obviously, the total number of equilibrium equivalence classes is at most $3^{n}$, and their union is  $\mathcal{U}^{*}$.

\subsection{Equilibrium Existence: Algorithm}


Below we explain how to recursively construct a strategy matrix which is a pure strategy Nash equilibrium for any parametric variation of the power allocation game.  

The recursion is only an example of a family of algorithms that can construct a pure strategy Nash equilibrium for \emph{any} parametric variation of the game, and thus proves equilibrium existence for the game -- it never seeks to exhaust the search of \emph{all} pure strategy Nash equilibria.

%


Let $q$ be the number of pairs in $\mathcal{R}_{A}$, and $\mathbf{q} = \{1,2, ..., q\}$ be the set of distinct labels for elements in $\mathcal{R}_{A}$. By an ordering map is meant an bijection $\gamma: \mathcal{R}_{A} \to \mathbf{q}$; any such map determines an ordering of $\mathcal{R}_{A}$ with $\{i,j\}$ the $\gamma(\{i,j\})$-th term in the ordering.

%
%

Let $z_{i}(k)$ be the $i$-th entry in $z(k)$ and $e_{k}$ be the $k$-th $n \times 1$ unit vector. Consider the recursion, \begin{align}
z(k) =  z(k-1) - \text{min}\{z_{i}(k-1), z_{j}(k-1)\}e_{k}
\end{align} where $k \in \mathbf{q}, z(k) \in \mathbb{R}^{1 \times n}$, $z(0) = p$, and $\{i,j\} = \gamma^{-1}(k-1)$. 

Let $e_{i}$ and $e_{j}$ be respectively the $i$-th and $j$-th $n \times 1$ unit vector. We let 
$U(k) = U(k-1) +$ \begin{align}  \text{min}\{z_{i}(k-1), z_{j}(k-1)\}(e_{i}e^{T}_{j} + e_{j}e^{T}_{i})
\end{align} where  $k \in \mathbf{q}$, $U(k) \in \mathbb{R}^{n \times n}$, $U(0) = \text{diagonal}\{z_{1}(q), z_{2}(q), ..., z_{n}(q)\}$, and $\{i,j\} = \gamma^{-1}(k-1)$.

Despite being a corner solution, $U$ may be the only kind of equilibria in many cases with non-generic parameters.  $U$ depends on the ordering map $\gamma$ we have chosen. Since there are $q!$ such maps, there are at least $q!$ pure strategy Nash equilibrium for the problem under consideration.  

In the sequel we will prove in Theorem~\ref{theorem/hob} that $U = U(q)$ is a pure strategy Nash equilibrium.

\begin{theorem}
\label{theorem/hob}
The game $\Gamma = \{\mathcal{C}, p, \mathcal{U}, \sigma_{i}, \tau_{i}, \mathcal{X}, \preccurlyeq\}$ always has pure strategy Nash equilibrium. 
\end{theorem}



\subsection{Equilibrium Existence: Proof}

The aim of this section is to prove Theorem~\ref{theorem/hob}. To do this it is useful to first establish certain property of the total power vector in Lemma~\ref{lemma/decomp} required for proving Theorem~\ref{theorem/hob}.


Our goal in Lemma~\ref{lemma/decomp} is to show that the existence of a pure strategy Nash equilibrium in the game is equivalent to \emph{a decomposability condition for the total power vector}. The decomposition makes use of an \emph{incidence  matrix} for the subgraph $\mathbb{G'}$ of all the adversary pairs of countries.

{\it Incidence Matrix for the Subgraph of All Adversary Pairs.} The $q$ adversary pairs make up a subgraph $\mathbb{G'}$. The incidence matrix of $\mathbb{G'}$ is $B = [b_{ik}]_{n \times q}$, where $b_{i}$ is its $i$-th row, and $b_{ik} = 1$ if country $i$ involves in the $k$-th ($0 \leq k \leq q$) adversary pair and $0$ otherwise. 

\begin{lemma}
\label{lemma/decomp}
The game $\Gamma = \{\mathcal{C}, p, \mathcal{U}, \sigma_{i}, \tau_{i}, \mathcal{X}, \preccurlyeq\}$ has a pure strategy Nash Equilibrium if the total power vector can be decomposed as 

\begin{align}
p^{T} = B d + c\end{align} where the following conditions are satisfied: 
\begin{enumerate}
\item $B$ is the $n \times q$ incidence matrix for the subgraph $\mathbb{G'}$ of the adversary relations. 
\item $d$ is an $q \times 1$ nonnegative-valued column vector, and $c$ is an $n \times 1$ nonnegative-valued column vector.
\item $\nexists \{i, j\} \in \mathcal{R}_{a}$, $c_{i} > 0 $ and $c_{j} > 0$.
\end{enumerate}
\end{lemma} 
\noindent {\bf Proof of Lemma~\ref{lemma/decomp}}: Suppose a decomposition of the total power vector that satisfies the three conditions exists. We then derive a strategy matrix $U$ as follows: 

\begin{enumerate}
\item Let allocations between countries in the $k$-th adversary pair be symmetric and equal to $d_{k}$, $u_{ij} = u_{ji} = d_{k}$ where $\gamma^{-1}(k) = \{i,j\} \in \mathcal{R}_{A}$ and $k \in \mathbf{q}$,
\item Let $u_{ii} = c_{i}$, $i \in \mathbf{n}$ 
\item As a consequence of the above, the allocations between friends are $0$, $u_{ij} = u_{ji} = 0$, $i \in \mathbf{n}$ and $j \in \mathcal{A}_{i}$.
\item We have stipulated before that $u_{ij} = u_{ji} = 0$, $i, j\in \mathbf{n}$ and $j \not\in \{i\} \cup \mathcal{A}_{i} \cup \mathcal{F}_{i}$
\end{enumerate}


$U$ is a valid strategy matrix because

$\sum_{j=1}^{n}u_{ij} = \sum_{j \in \mathcal{A}_{i}}u_{ij} +  u_{ii} =  b_{i}d + c_{i} = p_{i}$

No country $i$ with adversaries will unilaterally deviate from $u_{i}$, because it must fall into either case:

\begin{enumerate}
\item
$c_{i} = 0$. Here for any $j \in \mathcal{A}_{i}$, $u_{ij} = u_{ji}$, and that $u_{ii} = 0$. Thus $\sigma_{i}(U) = \tau_{i}(U) = p_{i}$, and $x_{i}(U) = \text{precarious}$. Thus, $i$ cannot deviate to make itself strictly better off, due to power deficiency.
\item
$c_{i} > 0$. Here it must be that $\forall j \in \mathcal{A}_{i}, c_{j} = 0$. Then $\forall j \in \mathcal{A}_{i}$, $x_{j}(U) = \text{unsafe}$ or $\text{precarious}$.  $\forall j \in \{i\} \cup \mathcal{F}_{i}$, $x_{j}(U) = \text{safe}$ or $\text{precarious}$.  

By Axioms 2 and 3, given any arbitrary $V \in \mathcal{U}$, $i$ must weakly prefer $U$ to $V$, i.e., it has achieved a best possible power allocation outcome $x(U)$, and therefore does not need to deviate. 
\end{enumerate}

Any country $i$ without adversaries will not unilaterally deviate from $u_{i}$, either, because the following holds for itself:

\begin{enumerate}

\item $p_{i} = u_{ii}$. 
\item $x_{i}(U) = \text{safe}$.
\item $x_{j}(U) = \text{safe}$ or $\text{precarious}$, $j \in \mathcal{A}_{i}$.

\end{enumerate} 
By Axioms 2 and 3, $i$ has also achieved a best possible power allocation outcome $x(U)$, and does not need to deviate. 

Therefore, if the total power vector can be decomposed as $p^{T} = Bd + c$ where those requirements hold, we can derive a strategy matrix that is pure strategy Nash equilibrium.  $\square$

Below we present the proof of Theorem~\ref{theorem/hob}, stating that the algorithm returns a strategy matrix that satisfies the decomposability condition in Lemma~\ref{lemma/decomp} and is thus a pure strategy Nash equilibrium. 

\noindent {\bf Proof of Theorem~\ref{theorem/hob}}: By the algorithm, we propose to decompose the total power vector as follows:

\begin{enumerate}
\item let $d_{k} = \min\{z_{i}(k-1), z_{j}(k-1)\}$ for $\gamma^{-1}(k) = \{i,j\}$, $k \in \mathbf{q}$
\item let $c_{i} = z_{i}(q)$, $i \in \mathbf{n}$.
\end{enumerate}

By the above, the first two conditions about the decomposition in Lemma~\ref{lemma/decomp} are thus proven -- $d$ is an $q \times 1$ nonnegative-valued column vector, and $c$ is an $n \times 1$ nonnegative-valued column vector. 

Now we aim to prove the third condition. By (1) of the algorithm, at the $k$-th recursion where $\gamma^{-1}(k) = \{i,j\}$ is traversed, $z_{i}(k)$ and $z_{j}(k)$ cannot be both positive. Also, $z_{i}(k)$ is non-increasing with $k$. Thus, $z_{i}(q)$ and $z_{j}(q)$ cannot be both positive, which means that $\nexists \{i, j\} \in \mathcal{R}_{a}$, $c_{i} > 0 $ and $c_{j} > 0$.



Lastly, the decomposition is valid, i.e., satisfying the total power constraint:
\begin{align}
b_{i}d + z_{i}(q) = p_{i}, i \in \mathbf{n}
\end{align}

A valid decomposition that satisfies three conditions in  Lemma~\ref{lemma/decomp} is thus derived based on the rules of the algorithm. Therefore, the game always has pure strategy Nash equilibrium. $\square$

%

\section{Conclusion}


Aside from equilibrium existence, the game has other theoretical results such as regarding the property of the equilibrium set and many unintuitive predictions that we will discuss in a fully fledged paper. Additionally, the game itself can be theoretically extended to study another fundamental question, which is how countries \emph{change the environment} toward where they can more easily pursue targets with power allocation, that is, through changing their own (1) relations  (2) or power (3) or both. Chapter 3 (``Theory of Relation Formation'') of \citep{yukedis} explores the first possibility theoretically, and we leave the study of the second and third possibilities for future extensions.

%
%
%
%


\section*{Acknowledgements}
The authors thank Man-Wah Cheung, Sander Heinsalu, Johannes H\"{o}rner, Ji Liu, Aniko \"{O}ry, Thomas Pogge, John Roemer, Nicholas Sambanis, Larry Samuelson, Milan Svolik, Weiyi Wu as well as many seminar participants for helpful discussions.

\bibliography{alliance}
\bibliographystyle{IEEEtran}

\renewcommand{\baselinestretch}{.4 }
\setlength{\parskip}{-0pt} \small\setlength{\baselineskip}{8pt}

\small

\renewcommand\refname{\large References}


\end{document}